\documentclass[aps,pre,floats,twocolumn,showpacs,superscriptaddress,unsortedaddress]{revtex4-1}
\usepackage{graphicx,epsfig}
\usepackage{graphics,dcolumn,bm,fleqn,epic,eepic,float}
\usepackage{amssymb,amsmath,multirow,color}
\usepackage{subfigure}
\usepackage{rotating}

\newcommand{\be}{\begin{equation}}
\newcommand{\ee}{\end{equation}}
\newcommand{\bea}{\begin{eqnarray}}
\newcommand{\eea}{\end{eqnarray}}
\newcommand{\bit}{\begin{itemize}}
\newcommand{\eit}{\end{itemize}}

\newcommand{\IN}{{\rm IN}}
\newcommand{\OUT}{{\rm OUT}}
\newcommand{\WCC}{{\rm WCC}}
\newcommand{\GWCC}{{\rm GWCC}}
\newcommand{\GSCC}{{\rm GSCC}}
\newcommand{\SCC}{{\rm SCC}}
\newcommand{\LSCC}{{\rm LSCC}}
\newtheorem{definition}{Definition}

\renewcommand{\emph}{\textbf}

\newcommand{\cm}[1]{}
\newcommand{\jt}[1]{}
\newcommand{\vl}[1]{}

\begin{document}
\title{Components in time-varying graphs}

\author{Vincenzo Nicosia}
\affiliation{Computer Laboratory, University of Cambridge, 15 JJ Thomson Av., Cambridge CB3 0FD, UK}
\affiliation{Laboratorio sui Sistemi Complessi, Scuola Superiore di Catania, Via Valdisavoia 9, 95123 Catania, Italy}
\author{John Tang}
\affiliation{Computer Laboratory, University of Cambridge, 15 JJ Thomson Av., Cambridge CB3 0FD, UK}
\author{Mirco Musolesi}
\affiliation{School of Computer Science, University of Birmingham, Edgbaston, Birmingham B15 2TT, UK}
\author{Giovanni Russo}
\affiliation{Dipartimento di Matematica ed Informatica, Universit\`a di Catania, Via S. Sofia, 64, 95123 Catania, Italy} 
\author{Cecilia Mascolo}
\affiliation{Computer Laboratory, University of Cambridge, 15 JJ Thomson Av., Cambridge CB3 0FD, UK}
\author{Vito Latora}
\affiliation{School of Mathematical Sciences, Queen Mary, University
  of London, E1 4NS, London, UK}
\affiliation{Dipartimento di Fisica e Astronomia, Universit\`a di Catania and INFN, Via S. Sofia, 64, 95123 Catania, Italy}
\affiliation{Laboratorio sui Sistemi Complessi, Scuola Superiore di Catania,Via Valdisavoia 9, 95123 Catania, Italy}

\date{\today}
\begin{abstract}
  Real complex systems are inherently time-varying. Thanks to new
  communication systems and novel technologies, it is today possible
  to produce and analyze social and biological networks with detailed
  information on the time of occurrence and duration of each link.
  However, standard graph metrics introduced so far in complex network
  theory are mainly suited for static graphs, i.e., graphs in which
  the links do not change over time, or graphs built from time-varying
  systems by aggregating all the links as if they were concurrent in
  time. In this paper, we extend the notion of connectedness, and the
  definitions of node and graph components, to the case of
  \textit{time-varying graphs}, which are represented as time-ordered
  sequences of graphs defined over a fixed set of nodes. We show that
  the problem of finding strongly connected components in a
  time-varying graph can be mapped into the problem of discovering the
  maximal-cliques in an opportunely constructed static graph, which we
  name the \textit{affine graph}. It is therefore an NP-complete
  problem. As a practical example, we have performed a temporal
  component analysis of time-varying graphs constructed from three
  data sets of human interactions. The results show that taking time
  into account in the definition of graph components allows to capture
  important features of real systems. In particular, we observe a
  large variability in the size of node temporal in- and
  out-components.  This is due to intrinsic fluctuations in the
  activity patterns of individuals, which cannot be detected by static
  graph analysis.
\end{abstract}
\maketitle

\textbf{ Time-varying graphs are a natural model for networked systems
  in which the relationships among nodes are intrinsically dynamic and
  fluctuate over time, where links appear and disappear at specific
  points in time and are often recurrent. Here we extend the concept
  of connectedness and the definitions of node and graph components to
  the case of time-varying graphs, we prove that finding strongly
  connected components in time-varying graphs is an NP-complete
  problem and we also report the results of component analysis
  performed on three real time-varying systems. This analysis confirms
  that the classical aggregate representations of networks evolving
  over time wash out most of the richness of the original systems. In
  particular, static graph erroneously flatten down fluctiations in
  the size of in- and out-componens of nodes, and tend to
  substantially overestimate the actual size of the connected
  components of the graph.}

\section{Introduction}

Complex network theory has proved to be a versatile framework to
represent and analyze biological, social and man-made complex
systems~\cite{boccaletti,arenas}.  Typically, a complex system is
inherently dynamic.  Social interactions and human activities are
intermittent~\cite{tadic_04,holme_network_2005,kleinberg_convergence_2008,kossinets_structure_2008},
the neighborhood of individuals moving over a geographic space evolves
over time~\cite{gonzalez,buscarino,barrat_crowd}, links appear and
disappear in the World Wide Web~\cite{tadic_01}, in patterns of
interactions among genes from microarray
experiments~\cite{caretta07,ahmed} and in functional brain
networks~\cite{valencia,fallani}.  In all these networks, time plays a
central role: links exist only for certain time periods, and are often
recurrent.  Despite this fact, most of the classic studies in complex
networks theory are based on the analysis of the topological
properties of {\em static graphs}. These are graphs in which the links
do not change over time, or graphs built from time-varying systems as
the result of the aggregation of all interactions, as if these were
all concurrent in time.  The evolution of linking patterns over time,
when considered, has been usually studied by creating a series of
graphs, each graph containing all the links appeared in a certain time
interval. Then, each standard graph metric has been evaluated for the
static graph obtained at each time window, and plotted as a function
of time~\cite{vicsek,clauset_persistence_2007}.  Today, thanks to
recent technological developments, for the first time we have the
opportunity to study large social and biological networks with precise
temporal information on the appearance, duration and frequency of
links among a set of nodes. Many other similar databases will be
produced in the near future, at an ever increasing rate.  These data
sets demand for new network measures and models that can take account
of the richness introduced by detailed temporal information.  Some
recent works have analyzed large interconnected systems with
fluctuating interactions~\cite{miritello,bianconi,grindrod_1}, and
some graph measures have been already extended to the case of graphs
in which connection patterns evolve over time~\cite{mucha}. More
recently, some works have also studied the onset of synchronization in
populations of agents interacting through time-evolving
topologies~\cite{diazguilera}. In previous
works~\cite{tang09,tang_wsns,tang_wowmom} it has been shown that a
static analysis of aggregated graphs is not able to capture the real
dynamic behavior and time correlations of complex networks evolving
over time. Since in this type of analysis the temporal ordering of
links is ignored, the number of available links at each time is
overestimated and, therefore, the actual lengths of walks and paths is
underestimated.  In particular, it has been found that edge causality
plays an important role on the dynamics of many processes that occur
on networks, such as disease
spreading~\cite{Moody,rochaPNAS,holme_immun}.  There have been several
attempts to incorporate temporal information in the description of
complex networks~\cite{Moody,kempe,cheng}, and different definitions
of time-varying graphs have been proposed so
far~\cite{kostakos,tang_pre10,pan}. At the same time, some basic
concepts of complex network theory, such as temporal
walks~\cite{grindrod_2,grindrod_3}, path length and
distance~\cite{kostakos,tang_pre10,pan}, centrality~\cite{tang_wsns}
have been extended to the case of temporal networks.

In this paper, we focus our attention on two important concepts in
graph theory, namely those of \textit{connectedness} and
\textit{connected components} of a graph, and we generalize them to
the case of time-varying graphs. These concepts have been thoroughly
used to study the reachability of pairs of nodes in static complex
networks~\cite{Broder00}, and to characterize the resilience of
networks to attacks~\cite{albert00}. Here we show that node
connectedness and connected components play a central role in
time-varying graphs, and can reveal interesting details on the real
structure of the network, that usually remain hidden to an analysis
based on static graphs.

The paper is organized as follows. In Section~\ref{sect:static}, we
briefly review the concepts of connectedness and components in static
graphs, while in Section~\ref{sect:tvg} we extend them to the case of
time-varying graphs. In particular, we define the temporal in- and
out-component of a given node, and we give the definition of weakly
and strongly connected components of a graph.  In
Section~\ref{sect:affine} we show that the problem of finding
components in a time-varying graph can be mapped into the
maximal-clique problem for an opportunely constructed graph, which we
call \textit{affine graph}. An affine graph is a static graph which
incorporates all the information on the temporal reachability of pairs
of nodes, and is a useful tool to analyze the components of the
corresponding time-varying graph. Thanks to this mapping, we also
prove that finding strongly connected components in time-varying
graphs is a NP-complete problem. Finally, in
Section~\ref{sect:results} we present the results of temporal
component analysis on time-varying graphs constructed from three
different data sets of human interactions.

\section{Components in static graphs}
\label{sect:static}
Let us consider a graph $G$ with $N$ nodes and $K$ links. From now on
we will refer to it as to a \textit{static graph}. We will consider
the case of undirected and directed static graphs separately. An {\em
  undirected static graph} $G$ can be represented by a symmetric
adjacency matrix, a $N \times N$ matrix $A$ whose each entry $a_{ij}$
is equal to one if and only if there is a link between $i$ and $j$,
and is equal to zero otherwise. In order to define graph components,
we need to introduce the concept of connectedness, first for pairs of
nodes, and then for the whole graph. Two nodes $i$ and $j$ of an
undirected graph $G$ are said to be {\em connected} if there exists a
path between $i$ and $j$. $G$ is said to be {\em connected} if all
pairs of nodes in $G$ are connected, otherwise it is said to be
unconnected or disconnected.  A {\em connected component} of $G$
associated to node $i$ is the maximal connected induced subgraph
containing $i$, i.e., the subgraph induced by all nodes
connected to node $i$. If an undirected graph is not connected, it is
always possible to find a partition of the graph into a set of
disjoint connected components. It is straightforward to prove that
this partition is unique.

A {\em directed static graph} $G$ is described in general by a
non-symmetric adjacency matrix, a $N \times N$ matrix $A$ whose each
entry $a_{ij}$ is equal to one if and only if there is a directed link
from $i$ to $j$, and is equal to zero otherwise. Defining
connectedness for pairs of nodes in a directed graph is more complex
than in an undirected graph, because a directed path may exist through
the network from vertex $i$ to vertex $j$, but this does not guarantee
that any path from $j$ to $i$ does actually exist.  Consequently, we
have two different definitions of connectedness between two nodes,
namely \textit{weak} and \textit{strong} connectedness. In particular,
we can define the {\em weakly} and the {\em strongly connected
  components} of a directed graph as follows~\cite{dorogovtsev01}. Two
nodes $i$ and $j$ of a directed graph $G$ are said {\em strongly
  connected} if there exists a path from $i$ to $j$ and a path from
$j$ to $i$.  A directed graph $G$ is said {\em strongly connected} if
all pairs of nodes $(i,j)$ are strongly connected.  A {\em strongly
  connected component} of $G$ associated to node $i$ is the maximal
strongly connected induced subgraph containing node $i$, i.e., the
subgraph induced by all nodes which are strongly connected to node
$i$.  A {\em weakly connected component} of $G$ is a component of its
{\em underlying undirected graph} $G^u$, which is obtained by removing
all directions in the edges of $G$. Two nodes $i$ and $j$ of $G$ are
\textit{weakly connected} if they are connected in $G^u$, and a
directed graph $G$ is said to be {\em weakly connected} if the
underlying undirected graph $G^u$ is connected. Hence, the components
of a directed graph can be of two different types, namely weakly and
strongly connected.  It is also useful to review the definitions of
components {\em associated to a node} of a directed graph. We have
four different definitions:

\begin{enumerate}

\item The {\em out-component of node $i$}, denoted as $\OUT(i)$, is
  the set of vertices $j$ such that there exists a directed path from
  $i$ to $j,\forall j$.
\item The {\em in-component of a node $i$}, denoted as $\IN(i)$, is
  the set of vertices $j$ such that there exists a directed path from
  $j$ to $i,\forall j$.
\item The {\em weakly connected component of a node $i$}, denoted as
  $\WCC(i)$, is the set of vertices $j$ such that there exists a path
  from $i$ to $j,\forall j$ in the underlying undirected graph
  $G^u$.
\item The {\em strongly connected component of a node $i$}, denoted as
  $\SCC(i)$, is the set of vertices $j$ such that there exists a
  directed path from $i$ to $j$ and also a directed path from $j$ to
  $i,\forall j$.
\end{enumerate}
We have implicitly used the last two concepts for the definitions of
weakly and strongly connected components of a directed graph given
above. In fact, the property of weakly and strongly connectedness
between two nodes is reflexive, symmetric and transitive, i.e., in
mathematical terms, it is an {\em equivalence relation}. Therefore, it
is possible to define weakly and strongly connected components of a
graph by means of the weakly and strongly connected components
associated to its nodes: a strongly (weakly) connected
component of a node is also a strongly (weakly) connected component of
the whole graph.
\begin{figure}[htbp]
  \centering
  \includegraphics[width=8.0cm,angle=0]{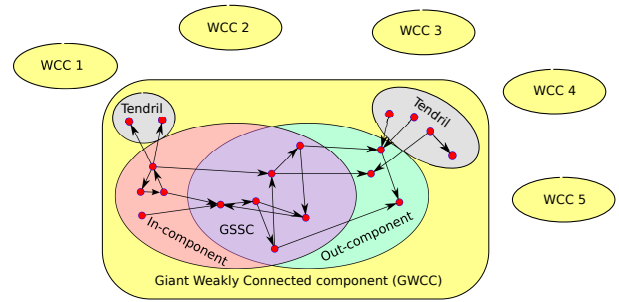}
  \caption{A directed graph can be partitioned into a set of disjoint
    weakly connected components (in yellow). Furthermore, each of
    these components has a rich internal structure, as shown for the
    $\GWCC$.}
  \label{fig06:directed}
\end{figure}
Conversely, the definitions of out-component and in-component of a
node are not based on {\em equivalence relations}.  In fact, the
symmetry property does not yield: $i \in \OUT(j)$ does not imply $j
\in \OUT(i)$. This means that out- and in-components can be associated
only to nodes, and cannot be directly extended to the entire graph. In
practice, we cannot partition a graph into a disjoint set of in- or
out-components, while it is possible to identify a partition of a
static graph into a disjoint set of weakly or strongly connected
components. However, the in- and out-components of the nodes of a
graph can be used to define the strongly connected components of the
graph. From the above definitions, we observe that $i\in \OUT(j)$ if
and only if $j\in\IN(i)$. Furthermore, we notice that $i$ and $j$ are
strongly connected if and only if $j\in \OUT(i)$ and, at the same
time, $i\in \OUT(j)$. Or equivalently, if and only if $j\in\OUT(i)$
and $j\in \IN(i)$. Therefore, the strongly connected component of node
$i$ is the intersection of $\IN(i)$ and $\OUT(i)$.

We are now ready to describe the rich interplay among the various
concepts of connectedness in a directed static graph in detail. In the
most general case, as shown in Fig.~\ref{fig06:directed}, a directed
graph can be decomposed into a set of disjoint weakly connected
components. In a large graph, one component will be larger than all
the others. This component is usually called the {\em giant weakly
  connected component} $\GWCC$ of the graph.
If we treat each link in the $\GWCC$ as bidirectional, then every node
in the $\GWCC$ is reachable from every other node in the $\GWCC$.  As
shown in Fig.~\ref{fig06:directed}, the $\GWCC$ contains the {\em
  giant strongly connected component} $\GSCC$, consisting of all nodes
reachable from each other following directed links. All the nodes
reachable from the $\GSCC$ are referred to as the {\em giant OUT
  component}, and the nodes from which the $\GSCC$ is reachable are
referred to as the {\em giant IN component}. The $\GSCC$ is the
intersection of the giant $\IN$- and $\OUT$-components. All nodes in
the $\GWCC$, but not in the $\IN$- and $\OUT$-components, are referred
to as ``tendrils''.

\section{Components in time-varying graphs}
\label{sect:tvg}

\begin{figure}[!htbp]
  \centering 
  \subfigure[]{
    \includegraphics[width=8cm]{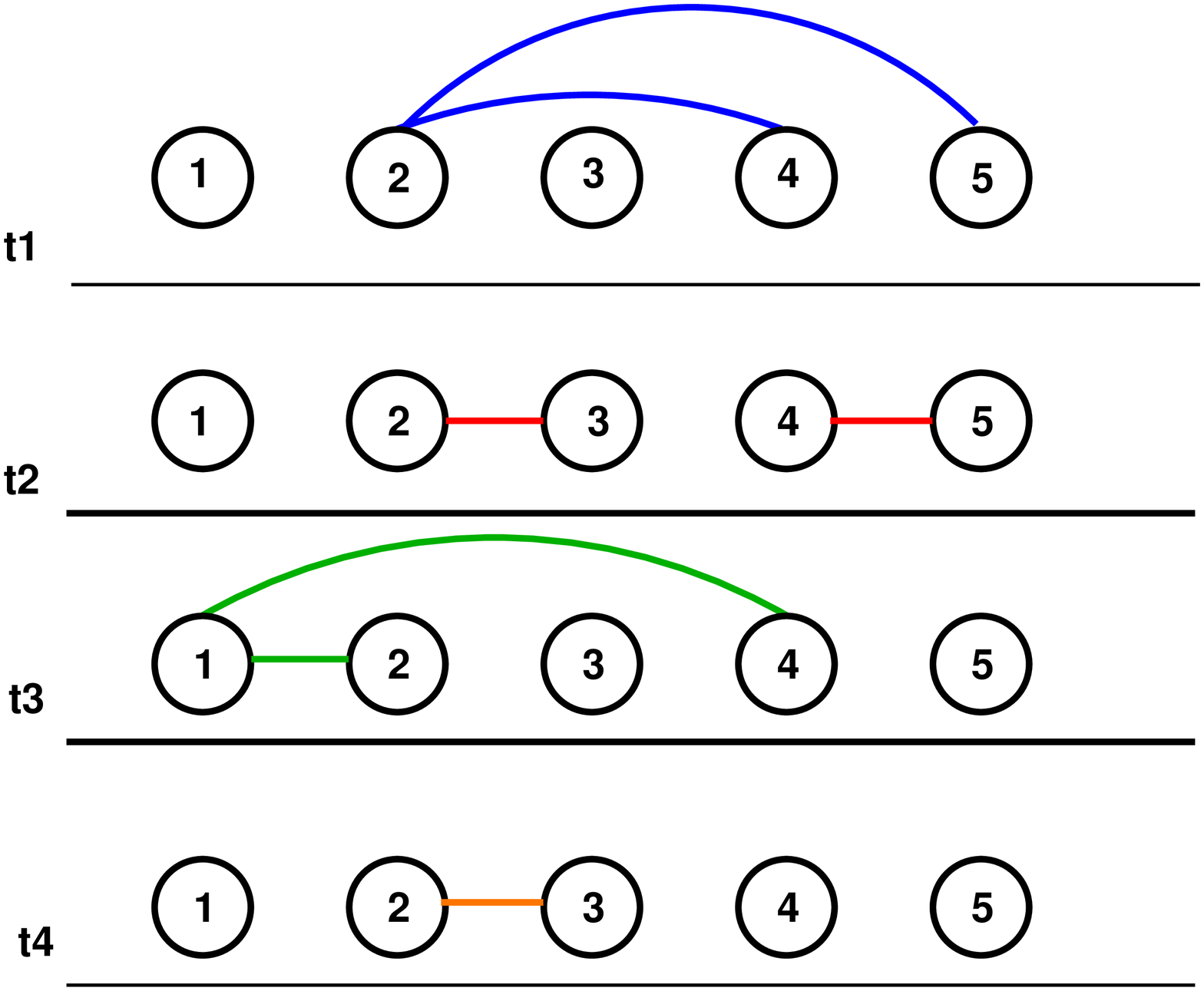}
  }\\\vspace{0.2cm}
  \subfigure[]{
    \includegraphics[scale=0.3]{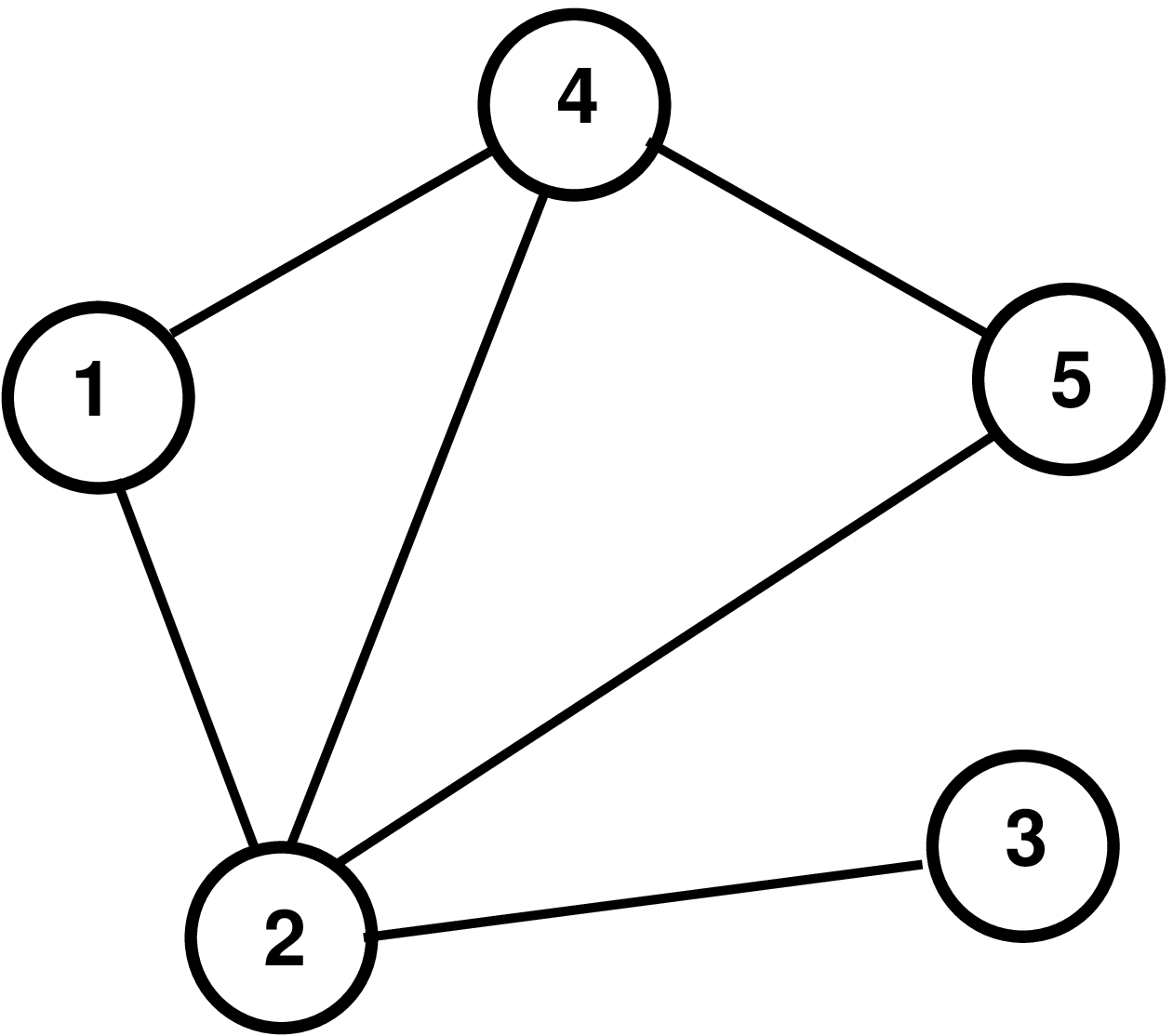}
  }
    \caption{A time-varying graph $\mathcal{ G}$ consisting of a
      sequence of $M=4$ graphs with $N=5$ nodes (panel a) and its
      corresponding aggregated static graph (panel b). The static
      representation of graphs discards time ordering of links and
      time correlations of paths. In the aggregated graph node $1$ and
      node $2$ are neighbors, but in the original time-varying graph
      they are directly connected only in one of the four graphs of
      the sequence, namely in $G_{t_3}$. Moreover, in the aggregated
      graph a path exists from node $1$ to node $5$ and vice-versa,
      while in the time-varying graph there exists a temporal path
      from $5$ to $1$ but there are no temporal paths from $1$ to
      $5$.}
  \label{fig:tvg}
\end{figure}

In this paper, we consider {\em time-varying graphs}, which are graphs
characterized by links that appear and disappear over time among a
fixed set of nodes.  A time-varying graph can be described as an
ordered sequence of graphs, i.e., an ordered set $\{
G_1,G_2,\ldots,G_M \}$ of $M$ graphs defined over $N$ nodes, where
each graph $G_m$ in the sequence represents the state of the network,
i.e., the configuration of links, at time $t_m$, where
$m=1,\ldots,M$. In this notation, the quantity $t_M -t_1$ is the
temporal length of the observation period. The graphs in the sequence
can be uniformly distributed over time, i.e., $t_{m+1} =t_m + \Delta
t, ~\forall m=1,\ldots,M-1$~\cite{tang_pre10}, or in general they can
correspond to any ordered sequence of times such that $t_1< t_2<
\ldots <t_M$~\cite{grindrod_2}. In a more compact notation, we denote
the graph sequence as $\mathcal{ G}\equiv \mathcal{G}_{[t_1,t_M]}$.
Each graph in the sequence can be either undirected or
directed. Consequently, the time-varying graph $\mathcal{G}$ can be
described by means of a time-dependent adjacency matrix
$A(t_{m}),~~m=1,\ldots,M$, where $a_{ij}(t_m)$ are the entries of the
adjacency matrix of the graph at time $t_m$. This matrix is in general
non-symmetric. If we discard the time-ordering of the links of a
time-varying graph $\mathcal{ G}$ and consider all links as concurrent
in time, we obtain its corresponding {\em aggregated static
  graph}. Different ways of constructing an aggregated graph from a
temporal network have been indeed proposed. It is possible to assign a
weight to each edge of the aggregated graph, according to the
frequency or the strength of the connection, and also to preserve the
directionality of edges of the original temporal system. However, all
these static representations fail to capture the temporal correlation
of links and, consequently, are not able to represent causality
relationships.  In panel a) of Fig.~\ref{fig:tvg} we report a simple
time-varying graph $\mathcal{G}_{[t_1,t_4]}$ with $N=5$ nodes and
$M=4$ edges and, in panel b), the corresponding aggregated static
graph (undirected and unweighted). It is worth noticing that the
aggregated graph discards most of the richness of the original
time-varying graph. For instance, three paths exist between node $1$
and node $5$ in the static aggregated graph, namely $1$-$4$-$5$,
$1$-$2$-$5$ and $1$-$2$-$4$-$5$, while in the time-varying graph there
is no temporal path from node $1$ to node $5$.
The problem of defining connectedness and components in time-varying
graphs looks more similar to the case of directed static graphs than
to the case of undirected static graphs. In fact, even if each graph
$G_m$, $m=1,\ldots,M$ in the sequence is undirected, the temporal
ordering of the graphs naturally introduces a directionality. For
instance, in the time-varying graph $\mathcal{G}_{[t_1,t_4]}$ reported
in Fig.~\ref{fig:tvg}, there exists a path connecting node $5$ to node
$1$ (i.e., the link $a_{52}$ at time $t_1$ and the link $a_{21}$ at
time $t_3$) but there is no path which connects node $1$ to node
$5$. An immediate consequence of this fact is that node $5$ can send a
message to node $1$ at time $t_1$, while node~$1$ cannot send a
message to node~$5$.

In order to define node connectedness for a time-varying graph, we
first need to introduce a mathematical definition of {\em
  reachability} for an ordered pair of nodes $i$ and $j$. We say that
$i$ can reach $j$ if $i$ can send a message to $j$ directly or through
a time-ordered sequence of contacts. In mathematical terms this
implies the existence of a walk connecting $i$ to $j$. In a
time-varying graph, a {\em walk}, also called {\em temporal walk},
from node $i$ to node $j$ is defined as a sequence of $L$ edges
$[(n_{r_0},n_{r_1}), (n_{r_1},n_{r_2}), \ldots,
  (n_{r_{L-1}},n_{r_L})]$, with $n_{r_0} \equiv i$, $n_{r_L} \equiv
j$, and an increasing sequence of times $t_{r_1} < t_{r_2} < \ldots <
t_{r_L}$ such that $a_{n_{r_{l-1}} , n_{r_l}}(r_l) \neq 0~~
l=1,\ldots,L$~\cite{grindrod_2,tang_pre10}.
A {\em path} (also called {\em temporal path}) of a time-varying graph
is a walk for which each node is visited at most once. For instance,
in the time-varying graph of Fig.~\ref{fig:tvg}, the sequence of edges
$[(5,2), (2,1)]$ together with the sequence of times $t_1,t_3$ is a
temporal path of the graph. This path starts at node $5$ at time $t_1$
and arrives at node $1$ at time $t_3$.  Given the definitions of
temporal walk and path, we can introduce the concepts of temporal
connectedness (in a weak and in a strong sense) for a pair of nodes.

A node $i$ of a time-varying graph $\mathcal{ G}_{[t_1, t_M]}$ is {\em
  temporally connected} to a node $j$ if there exists in $[t_1,t_M]$ a
temporal path going from $i$ to $j$. This relation is not symmetric:
if node $i$ is temporally connected to node $j$, in general node $j$
can be either temporally connected or disconnected to $i$.  In the
graph $\mathcal{G}_{[t_1,t_4]}$ of Fig.~\ref{fig:tvg}, node $5$ is
temporally connected to $1$ but node $1$ is not connected to node
$5$. For this reason, we introduce the definition of {\em strong
  connectedness}, which enforces symmetry:

\begin{definition}[\textbf{Strong connectedness}]
  \label{def:strong_connect}
  Two nodes $i$ and $j$ of a time-varying graph are {\em strongly
    connected} if $i$ is temporally connected to $j$ and also $j$ is
  temporally connected to $i$.
\end{definition}

Strong connectedness is a reflexive and symmetric relation, so that if
$i$ is strongly connected to $j$, then $j$ is strongly connected to
$i$. However, this definition of strong connectedness lacks
transitivity, and, therefore, it is not an equivalence relation. In
fact, if $i$ and $j$ are strongly connected and $j$ and $l$ are
strongly connected, nothing can be said, in general, about the
connectedness of $i$ and $l$. In the example shown in
Fig.~\ref{fig:tvg}, node $5$ and $2$ are strongly connected and also
$2$ and $1$ are strongly connected, but nodes $5$ and $1$ are not
strongly connected, since there exists no temporal path which connects
node~$1$ to node~$5$.  It is also possible to introduce the concept of
weak connectedness for a pair of nodes. Similarly to the case of
static directed graphs, given a time-varying graph $\mathcal{G}$, we
construct the underlying undirected time-varying graph
$\mathcal{G}^u$, which is obtained from $\mathcal{G}$ by discarding
the directionality of the links of all the graphs $\{G_m\}$, while
retaining their time ordering.

\begin{definition}[\textbf{Weak connectedness}]
  \label{def:weak_connect}
  Two nodes $i$ and $j$ of a time-varying graph are {\em weakly
    connected} if $i$ is temporally connected to $j$ and also $j$ is
  temporally connected to $i$ in the underlying undirected
  time-varying graph $\mathcal{G}^u$.
\end{definition}

Also weak connectedness is a reflexive and symmetric relation, but it
is not transitive. This definition of weak connectedness is quite
similar, but not identical, to that given for directed static
graphs. In fact, two nodes in $\mathcal{G}$ can be weakly connected
even if there is no temporal directed path which connects them, but
the temporal ordering of links breaks the transitivity so that if $i$
and $j$ are weakly connected and $j$ and $l$ are weakly connected,
then nothing can be said about the weak connectedness of $i$ and
$l$. Notice that the definitions of strong and weak connectedness
given above for time-varying graph are consistent with those given for
static graphs, so that if two nodes are strongly (weakly) connected in
a time-varying graph, then they are also strongly (weakly) connected
in the corresponding aggregated static graph. The vice-versa is
trivially not true. We are now ready to give the definitions of
components associated to a node of a time-varying graph $\mathcal{G}$:

\begin{enumerate}

\item The {\em temporal out-component of node $i$}, denoted as
  $\OUT_T(i)$, is the set of vertices which can be reached from $i$ in
  the time-varying graph $\mathcal{G}$.

\item The {\em temporal in-component of a node $i$}, denoted as
  $\IN_T(i)$, is the set of vertices from which $i$ can be reached in
  the time-varying graph $\mathcal{G}$.

\item The {\em temporal weakly connected component of a node $i$},
  denoted as $\WCC_T(i)$, is the set of vertices which $i$ can reach, 
  and from which $i$ can be reached, in the underlying undirected
  time-varying graph $\mathcal{G}^u$.

\item The {\em temporal strongly connected component of a node $i$},
  denoted as $\SCC_T(i)$, is the set of vertices from which vertex $i$
  can be reached, and which can be reached from $i$, in the
  time-varying graph $\mathcal{G}$.

\end{enumerate}

Differently from the case of directed static graphs, it is not
possible to define the strongly (weakly) connected components of a
time-varying graph starting from the definition of connectedness for
pairs of nodes. As we explained above, this is because the relation of
strongly (weakly) connectedness for couples of nodes is not an
equivalence relation. For this reason, we give the following
definition of strongly connected component of a time-varying graph:

\begin{definition}[\textbf{Strongly connected component}]
  \label{def:tv_scc}
  ~~~~~A set of nodes of a time-varying graph $\mathcal{G}$ is a
  temporal strongly connected component of $\mathcal{G}$ if each node
  of the set is strongly connected to all the other nodes in the set.
\end{definition}

Similarly, a set of nodes is a \textit{weakly connected component} if
each node in the set is weakly connected to all the other nodes in the
set.  The definitions of strongly and weakly connected components
enforce transitivity, but the check of strong (weak) connectedness has
to be directly performed for every couple of nodes. Let us suppose for
instance that we want to verify if the five nodes in the graph
$\mathcal{G}$ shown in Fig.~\ref{fig:tvg} form a strongly connected
component. In the static aggregated graph this check has $O(K)$
computational complexity, where $K$ is the total number of links in
the graph. In fact, we have only to check that $2$, $3$, $4$ and $5$
are connected to $1$, which can be done by a {\em depth first} visit
of the graph started at node $1$, since node connectedness is an
equivalence relation for static graphs, and a component of a node is
also a component for the whole graph. On the contrary, for a
time-varying graph we should check the connectedness of all the
possible couples of nodes, so that a procedure to verify that a set of
$N$ nodes forms a strongly connected component has computational
complexity $O(N^2)$ for every check, instead of $O(K)$.  Such a
difference in computational complexity is relevant for graphs obtained
from real world systems, which are usually sparse. Moreover, while
static directed graphs admit only one partition into strongly
connected components, for a time-varying graph there exists in general
more than one possible partition, as we shall see in the next section.

\section{The affine graph of a time-varying graph}
\label{sect:affine}
In this section we show that the problem of finding the strongly
connected components of a time-varying graph is equivalent to the
well-known problem of finding the maximal-cliques of an opportunely
constructed static graph~\cite{karp72}.  We call such a static graph
the \textit{affine graph} corresponding to the time-varying graph. It
is defined as follows:

\begin{definition}[\textbf{Affine graph of $\mathcal{ G}$}]
  ~~~~~~~~~~~~~~~~~~~~~~~~~~~~Given a time-varying graph $\mathcal{ G}\equiv
  \mathcal{G}_{[t_1,t_M]}$, the associated affine graph
  $G_{\mathcal{G}}$ is an undirected static graph with the same nodes
  as $\mathcal{G}$, and such that two nodes $i$ and $j$ are linked in
  $G_{\mathcal{G}}$ if $i$ and $j$ are strongly connected in
  $\mathcal{G}$.
\end{definition}
In practice, the affine graph of a time-varying graph can be obtained
by computing the temporal shortest paths between any two pairs of
nodes, and then adding a link between two nodes $i$ and $j$ of the
affine graph only if the temporal distance from $i$ to $j$ and the
temporal distance from $j$ to $i$ are both finite. Another method to
construct the affine graph is based on the usage of the
out-components of all the nodes. We start by considering the
out-component of the first node (let us say $i=1$) and then we check
if for each node $j \in OUT_T(i), j >i$ then also $i \in OUT_T(j)$. If
this is true, we add a link between $i$ and $j$ in the affine
graph. We then repeat this procedure for the second node, $i=2$, for
the third node, $i=3$ and so on. We obtain the affine graph by
iterating over the out-components of all the nodes. It is worth
noticing that the definition of affine graph is independent from the
particular definitions of reachability and temporal path given here,
since it is based only on the existence of mutual temporal paths
between pairs of nodes. Therefore, given any of the different
definitions of time-varying graph and the corresponding notions of
walk and path, it is always possible to construct the affine graph
associated to that particular representation of the system.

We also observe that the concept of affine graph is somehow related to
existing concepts of \textit{path graphs}, \textit{reachability
  graphs} and \textit{influence digraphs}~\cite{Moody,cheng}. In fact,
in a path graph (or in an influence digraph) a directed link exists
from node $i$ to node $j$ if there is a temporal path from $i$ to $j$,
i.e., if it is possible to reach $j$ starting from $i$. Therefore, we
can construct an affine graph from the path graph or from the
influence digraph associated to a time-varying graph, by replacing
each couple of reciprocated directed links with a single undirected
edge, and removing all the remaining links.

\begin{figure}
  \includegraphics[scale=0.23]{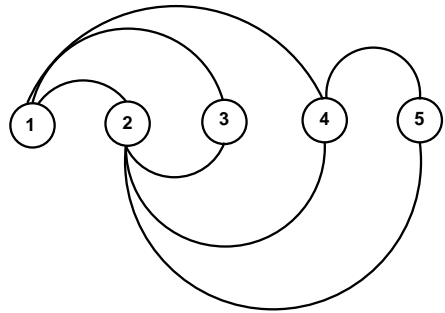}
  \caption{The affine graph $G_{\mathcal{G}}$ associated to the
    time-varying graph $\mathcal{G}$ reported in
    Fig.~\ref{fig:tvg}. The affine graph is static and undirected, and
    each of its maximal-cliques corresponds to a strongly connected
    component of the original time-varying graph $\mathcal{ G}$.}
  \label{fig:affine}
\end{figure}

In Fig.~\ref{fig:affine} we report the affine graph corresponding to
the time varying graph shown in Fig.~\ref{fig:tvg}.  In this graph,
node $1$ is directly connected to nodes $\{2,3,4\}$, since it is
temporally strongly connected to them in the time-varying
graph. Similarly, node $2$ is connected to nodes $\{1, 3, 4, 5\}$,
node $3$ is connected to $\{1,2\}$, node $4$ is connected to $\{1, 2,
5\}$ and node $5$ is connected to $\{2,4\}$. Hence, the affine graph
$G_{\mathcal{G}}$ has only 7 of the 10 possible links, each link
representing strong connectedness between two nodes.

We briefly report here some definitions about graph cliques. Given an
undirected static graph, a {\em clique} is a complete subgraph, i.e., a
subgraph in which all the nodes are directly linked to each other. A
\textit{maximal-clique} is a clique that is not included in any
larger clique, while a \textit{maximum-clique} is a {\em
  maximal-clique} whose size is equal to or larger than those of all the
other cliques~\cite{west}. 

By construction, a clique of the affine graph $G_{\mathcal{G}}$,
contains nodes which are strongly connected to each other, so that the
\textit{maximal-cliques} of the affine graph, i.e., all the cliques
which are not contained in any other clique, are temporal strongly
connected components ($\SCC_T$) of $\mathcal{G}$.  Similarly, all the
\textit{maximum-cliques} of the affine graph $G_{\mathcal{G}}$, i.e.,
its largest maximal-cliques, are the largest temporal strongly
connected components ($\LSCC_T$) of $\mathcal{G}$.  Therefore, the
affine graph can be used to study the connectedness of a time-varying
graph, and the properties of the strongly connected components of a
time-varying graphs can be obtained from known results about
maximal-cliques on static graphs. For instance, the problem of finding
a partition of $\mathcal{G}$ that contains the minimum number of
disjoint strongly connected components is equivalent to the
well--known problem of finding a partition of the corresponding affine
graph $G_{\mathcal{G}}$ in the smallest number of disjoint
maximal-cliques~\cite{karp72}. Unfortunately, this problem is known to
be NP--complete, and in practice can be exactly solved only for small
graphs. In the case of the affine graph in Fig.~\ref{fig:affine}, it
is possible to check by hand that there are only three possible
partitions of $G_{\mathcal{G}}$ into maximal-cliques, namely:
\begin{enumerate}
\item $\{1, 2, 3\} \bigcup \{4, 5\}$
\item $\{1, 2, 4\} \bigcup \{3\} \bigcup \{5\}$
\item $\{2, 4, 5\} \bigcup \{1, 3\}$
\end{enumerate}
Notice that the second partition contains two isolated nodes, which
are indeed degenerated maximal-cliques.  Therefore, the original
time-varying graph admits only two different partitions into a minimal
number of non-degenerated strongly connected components, namely into
two components containing at least two nodes each. One possible
partition of our network $\mathcal{G}_{[t_1,t_4]}$ is composed of the
components $\{1, 2, 3\}$ and $\{4, 5\}$, while the other partition
consists of $\{2, 4, 5\}$ and $\{1, 3\}$. If we discard the temporal
ordering of links, we obtain different results. In fact, the
aggregated static graph shown in Fig.~\ref{fig:tvg} has only one
connected component, which includes all the five nodes.

Other interesting results stem from the mapping into affine graphs and
from the following well known results for cliques in graphs.
\begin{enumerate}
\item
  Checking if a graph contains a clique of a given size $k$ has
  polynomial computational complexity, and precisely $O(N^k
  k^2)$~\cite{downey95}.
\item
  The {\em clique decision problem}, i.e., the problem of testing
  whether a graph contains a clique larger than a given size
  $\overline{k}$, is NP--complete~\cite{karp72}. Therefore, any
  algorithm that verifies if a time-varying graph has a strongly
  connected component whose size is larger than a fixed value
  $\overline{k}$, has exponential computational complexity.
\item Listing all the maximal-cliques of a graph has exponential
  computational complexity, namely $O(3^{N/3})$ on a graph with $N$
  nodes~\cite{moon65,bron73}. Consequently, finding all strongly
  connected components of a time-varying graph with $N$ nodes,
  requires an amount of time which exponentially grows with $N$.
\item 
  The problem of finding a maximum-clique for an undirected graph is
  known to be
  hard--to--approximate~\cite{feige91,arora98_1,arora98_2}, and an
  algorithm that finds maximum-cliques requires exponential time. This
  means that, if $P\neq NP$, there exists no polynomial algorithm
  to find approximated solutions for the maximum-clique problem. The
  best known algorithm works in $O(\sim 1.2^{N})$ for a graph with $N$
  nodes~\cite{tarjan77,robson86}.
\item
  The problem of determining if a graph can be partitioned into $K$
  different cliques is NP--complete, and consequently also the problem
  of finding the minimum number of cliques that cover a graph, known
  as the \textit{minimum clique cover}, is
  NP--complete~\cite{karp72}. This means that there exists no
  efficient algorithm to find a partition of a time-varying graph made
  by a set of disjoint strongly connected components. Moreover, there
  are in general more than one partition of a graph into
  maximal-cliques, so that a time-varying graph cannot be uniquely
  partitioned into a set of disjoint strongly connected components.
\end{enumerate}

The existence of a relation between the strongly connected components
of a time-varying graph and the maximal-cliques of its affine graph
implies that it is practically unfeasible to find all the strongly
connected components of large time-varying graphs. The problem can be
exactly solved only for relatively small networks, for which it is
computationally feasible to enumerate all the maximal-cliques of the
corresponding affine graphs. Even if, in many practical cases, it is
possible to find only the maximal-cliques up to a certain size
$\overline{k}$, we can still obtain some information about the maximum
value of $\overline{k}$ to be checked. First of all, in order to have
a clique of size $\overline{k}$ the graph should have at least
$\overline{k}$ nodes having at least $\overline{k}$ links. Moreover,
each clique of order $\overline{k}>3$ has exactly $\overline{k}
\choose 3 $ sub--cliques of order $3$, so that in order for a subgraph
to be a clique of order $\overline{k}$, the graph should have at least
$\overline{k} \choose 3$ triangles. This means that there is a
relation between the number of triangles of the affine graph and the
size of its maximum-cliques. In particular, the number of existing
triangles in the affine graph sets an upper bound for the size of the
largest admissible maximal-cliques of the graph.

\section{Results}
\label{sect:results}
In this section we extract and analyze node and graph components of
time-varying graphs constructed from three different data sets of
temporal social networks. The first is the data set of human
interactions produced by the Reality Mining Project~\cite{reality},
which was a large experiment devised and performed by researchers at
the MediaLab, Massachusetts Institute of Technology, US. One of the
aims of this experiment was to record the contacts among students,
staff and faculty members at the Massachusetts Institute of
Technology.  The resulting data set contains co--location information
among $100$ individuals during six months, from the end of June 2004
to the end of December 2004, sampled by means of Bluetooth--enabled
mobile phones at regular intervals of five minutes.  The second data
set contains all the contacts among $78$ people attending the INFOCOM
2006 conference~\cite{infocom}. Similarly to the Reality Mining data
set, also the INFOCOM 2006 dataset has been constructed using records
of scans made by Bluetooth devices provided to conference
participants, but covers a shorter period of time (only five days).
The third data set contains the records of communications among $\sim
100.000$ Facebook users living in Santa Barbara (CA), occurred over a
period of six months, from January 2009 to June
2009~\cite{facebook}. In the following we denote the Reality Mining,
INFOCOM and Facebook data sets as RM, IC and FB respectively.

When contacts are inferred from Bluetooth scans, we make the
reasonable assumption that two individuals are co--located, i.e., they
are at the same place, at a given time, if their respective devices
detect each other presence. In fact, Bluetooth devices are able to
detect similar devices within a very limited range, usually between
$5$ and $10$ meters. At each time $t$, a co--location graph can be
obtained by connecting through undirected links all the nodes which
are co--located at that time. For the RM and the IC data sets we
constructed several time-varying graphs, made of sequences of
co--location graphs obtained at regular intervals of $5$ minutes. For
the dataset of FB communication, the concept of co-location is not
meaningful, since the communication is usually asynchronous and
mediated by the Internet.  In this case, we construct a contact graph
for each hour, where a link between two nodes does exist if the nodes
have exchanged at least one message in the corresponding interval.  We
have chosen these three data sets because they represent different typical
aspects of temporal social networks. First of all, RM provides a good
example of recurrent face-to-face interactions, like those we
experience in our everyday life.  At the same time, IC is a relevant
instance of bursty face-to-face interactions, which are typical in
gatherings and meetings. Finally, FB is the prototypal model of
on-line communication, which is different from face-to-face
interaction because it does not require physical co--location to be
facilitated.

\begin{figure}[!htbp]
  \includegraphics[scale=0.35]{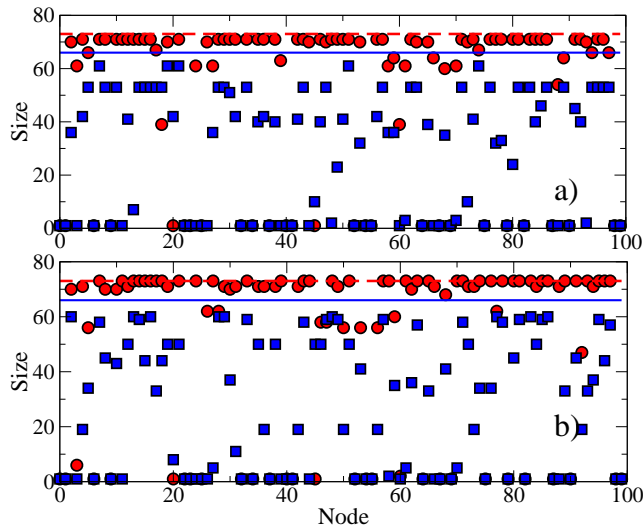}
  \caption{Size of the temporal in-component (panel a) and
    out-component (panel b) for each of the $N=100$ individuals during
    week $11$ of the RM data set. Red circles and blue
    squares correspond, respectively, to the beginning of the week
    (WB) and to end of the week (WE). For comparison, the size of the
    largest connected component of the corresponding aggregated static
    graph are reported as dashed red line (WB) and solid blue line
    (WE), respectively.}
  \label{fig:inout_comp}
\end{figure}

We start analyzing the distribution of the sizes of the node temporal
in- and out- components in RM, by considering time-varying graphs
obtained from the first half and the second half of a week.  In
particular, in Fig.~\ref{fig:inout_comp} we consider week 11, which
roughly corresponds to the beginning of the Fall term.
For each node, we report the size of its temporal in- and
out-component during the beginning of the week (WB), namely from
Monday 12:00am to Thursday 11:59am (red circles), and during the end
of the week (WE), namely from Thursday 12:00pm to Sunday 11:59pm (blue
squares). As shown in the figure, during WB almost all the nodes have
temporal in- and out-component of size $72$. Conversely, during WE, we
observe a wider distribution of the sizes of temporal in- and
out-components. In particular, in panel~(a) we notice a group of nodes
having an in-component of size $53$, another group whose in-component
contains around $40$ nodes, and other nodes with in-component of size
smaller than $30$.  Similarly, in panel~(b), there is a group of nodes
whose out-component contains around $60$ nodes, a second group of
nodes with out-component sizes between $40$ and $50$, and many other
nodes having out-component with less than $40$ nodes.  The observed
small variability in the size of node components during WB, is due to
the fact that students and faculty members have more opportunities to
meet and interact at lectures during WB. Even if not all students
attend the same classes, and not all professors teach to all the
students, there is a high probability that two individuals would be
connected by longer temporal paths.  Conversely, during WE, the
students usually meet other students in small groups, and they usually
do not meet professors and lecturers, except for the classes held on
Thursday afternoon and on Friday. As a result, the size of the in- and
out-components during WE exhibits large fluctuations from node to
node. Such fluctuations are lost in a static graph description, which
aggregates all the links independently of their time ordering. In
fact, the static aggregated graphs corresponding respectively to WB
and WE, have only one giant connected component which contain the
majority of the nodes, while the remaining nodes are isolated. As
comparison, the size of the giant component of the aggregated static
graphs for WB and WE are also reported in Fig.~\ref{fig:inout_comp},
respectively as dashed red line and solid blue line. Notice that the
static aggregated graph corresponding to a co--location time-varying
graph is intrinsically undirected. Therefore, the in- and
out-components of a node in this graph coincide and correspond to the
component to which the node belongs. Moreover, in a static aggregated
graph all the links (and consequently also all the paths) are always
available, so that all the nodes in the same connected component have
the same component size.  As a result, the variability in the node
connectedness of the time-varying network, which is evident from the
distribution of circles and squares in Fig.~\ref{fig:inout_comp}, is
flattened down in the aggregated static graphs. In the latter case,
all information about network connectedness is represented by a single
value, namely the size of the largest connected component, which does
not provide any information about the mutual reachability of two
generic nodes of such a component.  In particular, the size of the
giant connected component of the static aggregated graph is equal to
$74$ during WB and to $66$ during WE, despite the fact that in the
latter interval the majority of nodes has much smaller temporal in-
and out-components. Similar results are obtained for the temporal in-
and out-components of nodes in IC and FB. However, we do not report
here the corresponding figures for brevity.

We now analyze the strongly connected components of different
time-varying graphs obtained from the three data sets under
consideration. In particular, we focus on the Fall term of RM (namely
weeks from 10 to 19 in the original data set), the last four days of
IC (namely days 2 to 5), and the first twelve weeks of FB (from
January to the end of March 2009). In Table~\ref{table:table_RM} we
report the relevant structural properties of different affine graphs
constructed from RM. In particular, we consider the time-varying
graphs constructed in the first 24 hours (Monday) of ten consecutive
weeks (from week 10 to week 19), and the time-varying graphs
corresponding to the whole weeks. At the scale of one day, we observe
large fluctuations in the measured values.  The number of links $K$
ranges from $105$ in week 12 to $1485$ in week 15, while the number of
triangles $T$ is in the range $[307, 22096]$, with a mean value around
$10000$ and a standard deviation equal to $6932$. This variance is due
to the fact that, even if the daily activity of each individual is, on
average, almost periodic, in a particular day we can observe a
peculiar temporal pattern of connections. For instance, some students
decide to skip a class or the lessons are suspended for public
holidays. This is exactly what happens on week 12. Monday of week 12
is September 11$^{th}$ 2004, and corresponds to the \textit{Patriot
  Day}, a national holiday introduced in the US in October 2001,
designated in memory of the 2977 killed in the September 11$^{th}$,
2001 attacks.  Therefore, we observe the minimum connectivity and the
minimum number of triangles on week 12 of RM, because all teaching
activities were suspended, and students did not participate to lessons
as usual. At the scale of one week, instead, the number of links and
the number of triangles $T$ are much more stable over time: $K$ has a
mean value of $2717$ and a standard deviation of $342$, while $T$ has
a mean value around $62000$ and a standard deviation around
$12000$. For the graphs of Mondays, we find that the number $N_s$ and
the average size $\langle s \rangle$ of maximal cliques of the affine
graphs fluctuate over the ten weeks. In particular, we observe
relative smaller values of $N_s$ and $\langle s \rangle$ during
Mondays of weeks 10 to 14 than in weeks 15 to 19, which is probably
due to the relatively lower number of links and triangles. Conversely,
if we consider the size $S$ of the largest strongly connected
component (i.e., the largest maximal-clique of the affine graph), we
notice that it is not strongly correlated with $K$ and $T$. For
instance, the size of the largest strongly connected component found
at Monday of week 11 ($S=29$) is equal to that observed at Monday of
week 16. However, in week 11 the affine graph has a much smaller
number of links and triangles than in week 16. Moreover, on Monday of
week 14 we have a maximal-clique of size $27$, even if the number of
links and triangles is higher than on Monday of week 11.

\begin{table}
  \scriptsize
  \begin{tabular}{|c|c||c|c||c|d|c|c|c|c||c|}
    \hline 
    & & $K$ & $T$ & $ N_s$ & \multicolumn{1}{c|}{$\langle s \rangle$} 
    & $S$ & $N_S$ & $N_U$ & $N_I$ & $C$ \\ \hline
    \multirow{10}{4mm}{\begin{sideways}RM (Mondays)\end{sideways}}
    &10 & 646  & 4341  & 22 & 10.3 & 27 & 1  & 27  & 27 & 62\\ 
    &11 & 554  & 4414  & 15 & 9.1  & 29 & 1  & 29  & 29 & 54\\ 
    &12 & 105  & 307   & 11 & 4.1  & 13 & 1  & 13  & 13 & 22\\ 
    &13 & 772  & 8322  & 16 & 10.6 & 36 & 1  & 36  & 36 & 59\\ 
    &14 & 815  & 6481  & 20 & 12.7 & 27 & 1  & 27  & 27 & 62\\ 
    &15 & 1485 & 22096 & 23 & 23.7 & 44 & 1  & 44  & 44 & 67\\ 
    &16 & 1022 & 9033  & 22 & 16.5 & 29 & 1  & 29  & 29 & 70\\ 
    &17 & 1284 & 15572 & 19 & 22.3 & 38 & 1  & 38  & 38 & 67\\ 
    &18 & 1417 & 18430 & 16 & 20.7 & 44 & 1  & 44  & 44 & 67\\ 
    &19 & 1106 & 13531 & 13 & 20.9 & 38 & 2  & 42  & 34 & 60\\ 
    \hline\hline
    &  & $K$ & $T$ & $ N_s$ & \multicolumn{1}{c|}{$\langle s \rangle$} 
    & $S$ & $N_S$ & $N_U$ & $N_I$ &  C\\ \hline
    \multirow{10}{4mm}{\begin{sideways}RM (weeks)\end{sideways}}
    &10 & 2200 & 45428 & 10 & 44.0 & 61 & 1  & 61  & 61 & 69\\ 
    &11 & 2506 & 54500 & 12 & 46.8 & 64 & 1  & 64  & 64 & 75\\ 
    &12 & 2598 & 57913 & 12 & 43.5 & 66 & 1  & 66  & 66 & 77\\ 
    &13 & 2965 & 71561 &  9 & 62.5 & 69 & 1  & 69  & 69 & 79\\ 
    &14 & 2590 & 56826 & 15 & 39.3 & 64 & 1  & 64  & 64 & 79\\ 
    &15 & 3321 & 85348 &  9 & 54.7 & 74 & 1  & 74  & 74 & 85\\ 
    &16 & 2927 & 69452 &  9 & 53.2 & 70 & 1  & 70  & 70 & 80\\ 
    &17 & 2802 & 66247 & 10 & 57.9 & 69 & 1  & 69  & 69 & 77\\ 
    &18 & 2298 & 47429 & 12 & 40.0 & 61 & 2  & 62  & 60 & 73\\ 
    &19 & 2966 & 70963 & 13 & 53.8 & 69 & 3  & 71  & 68 & 81\\ 
    \hline
  \end{tabular}
  \caption{Structural properties of the affine graph corresponding to
    the time-varying graph of the first 24 hours of the week (Monday),
    and to the whole weeks in the Fall term of RM. We report the
    number of links ($K$), number of triangles ($T$), number of
    maximal cliques ($N_s$), average size of maximal cliques ($\langle
    s \rangle$), size of the largest maximal clique ($S$), number of
    largest maximal cliques ($N_S$), number of nodes in the union
    ($N_U$) and in the intersection ($N_I$) of all largest maximal
    cliques. The size of the giant component of the corresponding
    static aggregated graph ($C$) is reported in the rightmost
    column.}
  \label{table:table_RM}
\end{table}

\begin{table}
  \scriptsize
  \begin{tabular}{|c|c||c|c||c|d|c|c|c|c||c|}
    \hline 
    & & $K$ & $T$ & $ N_s$ & \multicolumn{1}{c|}{$\langle s \rangle$}  
    & $S$ & $N_S$ &  $N_U$ & $N_I$ & $C$ \\ \hline
    \multirow{15}{4mm}{\begin{sideways}IC (hours)\end{sideways}} &
    6:00 & 376 & 2061 & 17 & 8.1 & 19 & 2 & 20 & 18 & 44 
    \\
    &7:00 & 793 & 7514 & 16 & 15.8 & 29 & 1 & 29 & 29 & 50 
    \\
    &8:00 & 1922 & 35472 & 13 & 35.8 & 56 & 2 & 57 & 55 & 69 
    \\
    &9:00 & 1825 & 35598 & 4 & 29.5 & 60 & 1 & 60 & 60 & 61 
    \\
    &10:00 & 2252 & 48459 & 6 & 50.8 & 64 & 1 & 64 & 64 & 68
    \\
    &11:00 & 2142 & 45510 & 3 & 43.0 & 65 & 1 & 65 & 65 & 66 
    \\
    &12:00 & 2346 & 52394 & 1 & 69.0 & 69 & 1 & 69 & 69 & 69 
    \\
    &13:00 & 2621 & 61270 & 2 & 69.0 & 72 & 1 & 72 & 72 & 73
    \\
    &14:00 & 2389 & 53297 & 2 & 56.5 & 69 & 1 & 69 & 69 & 70 
    \\
    &15:00 & 2330 & 51382 & 4 & 63.5 & 66 & 1 & 66 & 66 & 69 
    \\
    &16:00 & 1991 & 40153 & 5 & 46.6 & 58 & 3 & 60 & 56 & 65 
    \\
    &17:00 & 1862 & 36165 & 5 & 44.2 & 58 & 1 & 58 & 58 & 62 
    \\
    &18:00 & 1595 & 27979 & 6 & 39.3 & 51 & 1 & 51 & 51 & 58 
    \\
    &19:00 & 1177 & 18424 & 2 & 25.5 & 49 & 1 & 49 & 49 & 49 
    \\
    &20:00 & 1110 & 16393 & 8 & 16.0 & 46 & 1 & 46 & 46 & 49 
    \\ \hline\hline
    & & $K$ & $T$ & $ N_s$ & \multicolumn{1}{c|}{$\langle s \rangle$} 
    & $S$ & $N_S$ & $N_U$ & $N_I$ & $C$ \\ \hline
    \multirow{4}{4mm}{\begin{sideways}IC (days)\end{sideways}} &
    2 & 2920 & 72638 & 4 & 55.8 & 75 & 1 & 75 & 75 & 78 
    \\ 
    &3 & 2776 & 67435 & 3 & 50.3 & 74 & 1 & 74 & 74 & 76 
    \\ 
    &4 & 2225 & 47287 & 5 & 44.0 & 63 & 1 & 63 & 63 & 68
    \\ 
    &5 & 1602 & 27617 & 9 & 31.7 & 53 & 1 & 53 & 53 & 60 
    \\ 
    \hline
    
  \end{tabular}
  \caption{Structural properties of the affine graphs corresponding to
    time-varying graphs of different hours of the third day and of
    each of the four days of IC. The graph corresponding to each hour
    includes all the contacts recorded in that hour, so that, for
    instance, the first graph is constructed from the interactions
    observed from $6:00$ to $6:59$. Legend as in
    Table~\ref{table:table_RM}.}
  \label{table:table_IC}
\end{table}

At the scale of a week, we still observe relevant fluctuations of
average size $\langle s \rangle$, in the range $[39.3, 62.5]$, but the
number of strongly connected components is stable around $N_s=10$,
with a maximum of $N_S=15$ in week 14. These results confirm that the
size of the largest strongly connected component of a time-varying
graph is determined by the actual configuration of links and triangles
of the corresponding affine graph, and not only by their relative
number, and in general depends on the temporal scale at which we
observe the system.

Examining the results reported in Table~\ref{table:table_RM}, we
notice that the affine graphs for Mondays of all the weeks of RM admit
a single $\LSCC_T$, except in week 19 where two $\LSCC_T$s of size
$S=38$ emerge. For this reason, we also looked at the number of nodes
$N_U$ which participate to \textit{at least one} $\LSCC_T$, and at the
number $N_I$ of nodes which participate to \textit{all}
$\LSCC_T$s. These numbers correspond, respectively, to the number of
nodes found in the union and in the intersection of all $\LSCC_T$s. An
interesting result is that $N_I=34$ on Monday of week 19, so that $34$
nodes participate to both maximal $42$-node cliques. These $34$ nodes
play a very important role in the structure of the network. If we
remove just one of them, then the resulting affine graph does not have
a clique of size $42$ any more, and consequently the size of the
$\LSCC_T$ of the remaining time-varying graph is smaller than $42$. At
the same time, removing all these $N_I$ nodes will cause a significant
reduction in the size of $\LSCC_T$s, in the number of triangles of the
affine graph and, consequently, in the number of $\SCC_T$s. The nodes
that participate to at least one $\LSCC_T$ are important for the
diffusion of information throughout time-varying graphs. In fact, if a
message is passed to one of these $N_U$ nodes, then there is a high
probability that the message could be delivered to all the other $N_U
-1$ nodes in the union of all $\LSCC_T$.  Similar results are observed
also at the scale of a week, where almost all graphs have just one
$\LSCC_T$, except for week 18 and 19 where we found $N_S=2$ and
$N_S=3$, respectively.

In the rightmost column of Table~\ref{table:table_RM} we report, for
each time-varying graph, the size $C$ of the giant component of the
corresponding static aggregated graph. At the scale of one day, the
value of $C$ is always much larger than $S$, as a consequence of the
fact that the static representation of the time-varying graph
systematically overestimates node connectedness and paths
availability. Conversely, at the scale of one week the size of the
$\LSCC_T$ is comparable with the size of the largest strongly
connected component of the aggregated graph. In fact, when we consider
longer time periods, longer temporal paths appear and, consequently,
there is a higher probability that a large number of pairs of nodes
become mutually reachable. The affine graphs of entire weeks of RM are
almost complete graphs and look similar to each other, so that we
observe a relatively high correlation between $S$ and $C$ at the scale
of one week, even if there is no correlation between $S$ and $C$ at
the scale of a single day. For instance, on Monday of week 16 we
observe the maximum value of $C$, namely $C=70$, while the
corresponding time-varying graph has a largest strongly connected
component of size $S=29$, which is relatively small compared to the
other weeks.  Conversely, on Monday of week 13 we observe a relatively
small giant component, with $C=59$ nodes, while the size of the
largest strongly connected component at the same day is $S=36$.

In Table~\ref{table:table_IC} we consider the time-varying graphs
constructed at different hours of the day 3 of IC, and those obtained
for each of the four days of the conference. In this case, the system
at the scale of one day is saturated: there are just a few strongly
connected components and only one $\LSCC_T$, which roughly correspond
with the largest connected component of the aggregated graph. In fact,
the dynamics of interactions between participants at a conference
differs a lot from the recurrent daily interactions which are typical
in RM. Usually, during a single day of conference the participants
attend different plenary sessions, and consequently they have many
opportunities to be co--located with a large number of other
participants for a relatively long period (at least a couple of
hours). Therefore, all the participants to the same session usually
form a single strongly connected component. However, if we observe the
system at the scale of one hour, we notice that there are relevant
fluctuations in the number of links and triangles, and in the number
and size of strongly connected components. In the graphs constructed
from contacts recorded before 8:00 we observe a relatively smaller
number of links and triangles and smaller values of $S$. The number of
links remains almost stable until 17:00, while $S$ steadily increases
up to $S=72$ at 13:00 (lunch time) when the size of the $\LSCC_T$
reaches a maximum. Then, both the number of triangles and the size of
the $\LSCC_T$ decrease, and the graph at 20:00 has $S=46$ and
$T=16393$. Also at the scale of one hour we observe a clear
correlation between $S$ and $C$, $N_S=1$ and the average size $\langle
s \rangle$ of $\SCC_T$s is comparable with the size of the largest
strongly connected component.

\begin{table}[!htbp]
  \scriptsize
  \begin{tabular}{|c|c||c|c||c|d|c|c|c|c||c|}
    \hline
    & & $K$ & $T$ & $ N_s$ & \multicolumn{1}{c|}{$\langle s \rangle$} 
    & $S$ & $N_S$ & $N_U$ & $N_I$ &  C\\ \hline
    \multirow{12}{4mm}{\begin{sideways}FB (weeks)\end{sideways}} &
    1 & 43491 & 52284 & 6445 & 4.2 & 22 & 1 & 22 & 22 & 12000
    \\ 
    & 2 & 48404 & 50364 & 7583 & 4.1 & 20 & 1 & 20 & 20 & 13998 
    \\ 
    & 3 & 43400 & 32655 & 6922 & 3.8 & 16 & 1 & 16 & 16 & 12773 
    \\ 
    & 4 & 60853 & 102066 & 9377 & 4.3 & 41 & 1 & 41 & 41 & 17933
    \\ 
    & 5 & 65703 & 97631 & 10182 & 4.4 & 23 & 7 & 29 & 15 & 19973
    \\ 
    & 6 & 70282 & 117530 & 10646 & 4.5 & 27 & 1 & 27 & 27 & 20976
    \\ 
    & 7 & 60666 & 78769 & 9335 & 4.2 & 28 & 1 & 28 & 28 & 18537
    \\ 
    & 8 & 73772 & 188457 & 10709 & 4.7 & 46 & 3 & 48 & 45 & 20256
    \\ 
    & 9 & 79645 & 200061 & 11475 & 4.8 & 38 & 1 & 38 & 38 & 21990
    \\ 
    & 10 & 66849 & 85373 & 10595 & 4.3 & 18 & 1 & 18 & 18 & 20425
    \\ 
    & 11 & 55040 & 56615 & 8753 & 4.0 & 27 & 2 & 28 & 26 & 18266
    \\ 
    & 12 & 51418 & 59858 & 7735 & 4.1 & 28 & 2 & 29 & 27 & 15667
    \\ 
    \hline\hline
    & & $K$ & $T$ & $ N_s$ & \multicolumn{1}{c|}{$\langle s \rangle$} 
    & $S$ & $N_S$ & $N_U$ & $N_I$ &  C\\ \hline
    \multirow{12}{4mm}{\begin{sideways}FB (adj.weeks)\end{sideways}} &
    1 & 197818 & 2542462 & 48782 & 29.8 &112 & 1 & 112 & 112 & 32071 
    \\ 
    &2 & 170830 & 990149 & 17325& 11.8 & 58 & 1 & 58 & 58 & 33114 
    \\ 
    &3 & 211523 & 1936214 & 29947 & 17.5 & 89 &1 & 89 & 89 & 36820
    \\ 
    &4 & 327351 & 135246 & 27432 & 19.3 & 88 &1 & 88 & 88 & 37455
    \\ 
    &5 & 419563 & 8977497 & 1355049 & 42.5 & 119 & 1 & 119 & 119 & 44464 
    \\ 
    &6 & 345943 & 4805124 & 392966 & 31.0 & 91& 1 & 91 & 91 & 44117 
    \\ 
    &7 & 346856 & 5527618 & 149156 & 23.0 & 145 & 1 & 145 & 145 & 44107
    \\ 
    &8 & 313564 & 3176483 & 195238 & 25.2 & 138 & 1 & 138 & 138 & 44886
    \\ 
    &9 & 493502 & 14700802 & 1160920 & 66.7 & 149 & 1 & 149 & 149 & 46790 
    \\ 
    &10 & 282759 & 2966668 & 154016 & 21.5 & 87 & 1 & 87 & 87 & 43693 
    \\ 
    &11 & 217796 & 2277735 & 29077  & 13.2 & 103 & 2 & 104 & 102 & 39950
    \\ 
    \hline
  \end{tabular}
  \caption{Structural properties of the affine graphs corresponding to
    the time-varying graphs of weeks (upper rows) and pairs of
    adjacent weeks (lower rows) of FB data set. Legend as in
    Table~\ref{table:table_RM}.}
  \label{table:table_FB}
\end{table}

Finally, in Table~\ref{table:table_FB} we report the structural
properties of the affine graphs constructed from FB. We considered 12
time-varying graphs at the scale of a week and 11 time-varying graphs
at the scale of 14 days, i.e. couples of subsequent weeks. In this
case the number of nodes of the affine graphs was quite large, so we
used HPC facilities provided by the Edinburgh Parallel Computing
Center (EPCC), in order to run a parallelized version of the algorithm
described in~\cite{robson86} on 128 processors. At the scale of a
week, we observe significant fluctuations in the structural properties
of the affine graphs corresponding to distinct weeks: there is a high
variability in the number of links ($K$ is in the range
$[43400,79645]$) and in the number of triangles ($T$ is in the range
$[32655,200061]$). However, the size of the $\LSCC_T$ is relatively
small, compared to the number of nodes in the graph. In fact, $S$
remains in the range $[16,41]$, which is quite surprising for a graph
with more than $100000$ nodes. At the same time, there is no
correlation at all neither between $S$ and $C$ nor between $S$ and
$T$. A similar result is obtained when considering pairs of adjacent
weeks: in this case, the size of the $\LSCC_T$ is slightly larger (the
maximum is observed for the 9$^{th}$ pair of weeks, when $S=149$), but
$S$ is still two orders of magnitude smaller than $C$. The case of FB
is emblematic of the implicit weaknesses of a component analysis based
on static aggregated graphs. According to traditional static graphs
theory, the giant component of a network constructed from two adjacent
weeks of Facebook communication contains at least $32\%$ of the 100000
nodes, while a temporal analysis reveals that less than $0.15\%$ of
the nodes are indeed mutually reachable.

\section{Conclusions}
\label{sec:conclusion}

Conventional definitions of connectedness and components proposed so
far have only considered aggregate, static topologies, neglecting
important temporal information such as time order, duration and
frequency of links.  In this work we have extended the concepts of
connectedness to the case of time-varying graphs, and we have
introduced definitions of node and graph components which take into
account duration, times of appearance and temporal correlations of
links. The proposed temporal measures are able to capture variations
and fluctuations in the linking patterns, typical of many real social
and biological systems. As a first application, we have studied three
databases of human contacts, showing that variations in the pattern of
connections among nodes produce relevant differences in the size and
number of temporal strongly connected components. We have pointed out
the important role played by nodes that belong to different strongly
connected components at the same time, and we have also analyzed how
the size of temporal strongly connected components depend on the
observation interval. We hope that our formalism will be useful to
analyze other data sets of time-varying networks that will be
available in the near future, and to better characterize dynamical
processes that take place on these networks, such as diffusion of
information and spreading of diseases.

\begin{acknowledgments}
This work was partially carried out under the HPC-Europa2 project
(project number: 228398) with the support of the European Commission
Capacities Area-Research Infrastructures Initiative. This work made
also use of the facilities of HECToR, the UK’s national high
performance computing service, which is provided by UoE HPCx Ltd at
the University of Edinburgh, Cray Inc and NAG Ltd, and funded by the
Office of Science and Technology through EPSRC’s High End Computing
Programme. This work has been partially supported by the EPSRC Project
MOLTEN (EP/1017321/1).
\end{acknowledgments}

\end{document}